\documentclass[sigconf]{acmart}

\usepackage{textcomp}
\usepackage{stfloats}
\usepackage{url}
\usepackage{verbatim}
\usepackage{amsthm}
\usepackage{subfigure}
\usepackage{makecell}
\usepackage{multirow}
\usepackage{enumitem}
\usepackage{subcaption}

\usepackage{color, colortbl}

\definecolor{Gray}{rgb}{0.9, 0.9, 0.9}

\usepackage[capitalize,noabbrev]{cleveref}

\usepackage[ruled,linesnumbered]{algorithm2e}

\usepackage[inkscapelatex=false]{svg}

\AtBeginDocument{%
  }

\copyrightyear{2025}
\acmYear{2025}
\setcopyright{acmlicensed}\acmConference[MM '25]{Proceedings of the 33rd ACM International Conference on Multimedia}{October 27--31, 2025}{Dublin, Ireland}
\acmBooktitle{Proceedings of the 33rd ACM International Conference on Multimedia (MM '25), October 27--31, 2025, Dublin, Ireland}
\acmDOI{10.1145/3746027.3755781}
\acmISBN{979-8-4007-2035-2/2025/10}

\begin{document}

\title[FastMMRec]{The Best is Yet to Come: Graph Convolution in the Testing Phase for Multimodal Recommendation}


\author{Jinfeng Xu}
\email{jinfeng@connect.hku.hk}
\affiliation{%
  \institution{The University of Hong Kong}
  \city{HongKong SAR}
  \country{China}}

\author{Zheyu Chen}
\email{zheyu.chen@connect.polyu.hk}
\affiliation{%
  \institution{The Hong Kong Polytechnic University}
  \city{HongKong SAR}
  \country{China}}

\author{Shuo Yang}
\email{shuoyang.ee@gmail.com}
\affiliation{%
  \institution{The University of Hong Kong}
  \city{HongKong SAR}
  \country{China}}

\author{Jinze Li}
\email{lijinze-hku@connect.hku.hk}
\affiliation{%
  \institution{The University of Hong Kong}
  \city{HongKong SAR}
  \country{China}}


\author{Edith C. H. Ngai}
\authornote{*Corresponding authors}
\email{chngai@eee.hku.hk}
\affiliation{%
  \institution{The University of Hong Kong}
  \city{HongKong SAR}
  \country{China}}

\renewcommand{\shortauthors}{Xu et al.}

\begin{abstract}
The efficiency and scalability of graph convolution networks (GCNs) in training recommender systems remain critical challenges, hindering their practical deployment in real-world scenarios. In the multimodal recommendation (MMRec) field, training GCNs requires more expensive time and space costs and exacerbates the gap between different modalities, resulting in sub-optimal recommendation accuracy. This paper critically points out the inherent challenges associated with adopting GCNs during the training phase in MMRec, revealing that GCNs inevitably create unhelpful and even harmful pairs during model optimization and isolate different modalities. To this end, we propose FastMMRec, a highly efficient multimodal recommendation framework that deploys graph convolutions exclusively during the testing phase, bypassing their use in training. We demonstrate that adopting GCNs solely in the testing phase significantly improves the model's efficiency and scalability while alleviating the modality isolation problem often caused by using GCNs during the training phase. We conduct extensive experiments on three public datasets, consistently demonstrating the performance superiority of FastMMRec over competitive baselines while achieving efficiency and scalability.
\end{abstract}



\maketitle

\section{Introduction}
Multimodal recommendation (MMRec) plays a pivotal role in e-commerce and content-sharing platforms, encompassing a amount of web multimedia content, including descriptions and images \cite{xu2025mentor,zhou2023tale}. Such capabilities allow them to discern users' preferences across different modalities accurately. Several recent studies incorporate multimodal content into multimedia recommendation systems. For example, VBPR \cite{he2016vbpr} expands the matrix decomposition framework to accommodate item modality features. ACF \cite{chen2017attentive} innovates with a hierarchically structured attention network designed to discern user preferences at the component level. Improving the performance of recommendation models with Graph Convolutional Networks (GCNs) has gained widespread attention \cite{xu2025nlgcl,xu2024aligngroup,he2020lightgcn,xu2024fourierkangcf,xu2024improving}. More recently, models such as MMGCN \cite{wei2019mmgcn} and GRCN \cite{wei2020graph} employ GCNs to integrate modality information into message-passing processes, thereby enhancing the inference of user and item representations. To further explore the rich multimodal information of items, LATTICE \cite{zhang2021mining} and FREEDOM \cite{zhou2023tale} construct item-item graphs to aggregate semantically similar items. Despite the notable advancements in graph-based MMRec models, they encounter fundamental challenges \cite{hamilton2017inductive,ying2018graph} related to efficiency and scalability. These challenges stem primarily from the computationally intensive message-passing mechanisms of graph convolution, which are integral to the prevailing training paradigms of graph-based recommendation systems. The deployment of these models on large-scale graphs in real-world applications further exacerbates these challenges, as both time and computational complexity increase exponentially with the growing number of users and items. To make models scalable for real-world deployment, research focuses on two perspectives:
\begin{itemize}[leftmargin=*]
    \item \textbf{Perspective 1:} Extensive studies have been devoted to designing GCNs with complexity that is approximately linear or sublinear for the size of the data \cite{chen2024macro,wang2024fast}, with a wide range of research focusing on sampling methods. Sampling-based methods lower the computation and memory requirements of GCNs by using a mini-batch training strategy on GCNs, which samples a limited number of neighbors for target nodes in a node-wise \cite{chen2018stochastic,cong2020minimal,hamilton2017inductive}, layer-wise \cite{chen2018fastgcn,zou2019layer}, or subgraph-wise \cite{zenggraphsaint,zhang2024linear} manner. However, sampling-based methods inevitably omit a large number of neighbors for aggregation, resulting in large random errors. 
    \item \textbf{Perspective 2:} Extensive studies show that simple MLPs as the initialization of graph model \cite{hanmlpinit2023,yang2023graph} or trained with contrastive learning \cite{hu2021graph,yang2023graph,yang2024your}, knowledge distillation \cite{zhanggraph} demonstrate competitive performance compared with GCN models as long as they share an equivalent weight space.
\end{itemize}
\textbf{Perspective 1} verifies the importance of a complete graph structure for GCN, and \textbf{Perspective 2} explores viable alternatives to GCNs. Therefore, we naturally raise a meaningful and significant question:
\begin{gather*}
    \textbf{What do GCNs actually do during Training?}
\end{gather*}

To answer this problem, we analyze the impact of GCNs on the model during training in Section~\ref{sec: Investigation}. Then we point out two major challenges posed by employing GCNs in the training phase, including \textbf{GCNs inevitably create unhelpful or even harmful positive and negative pairs during model optimization.} and \textbf{C2. GCNs isolate different modalities, resulting in sub-optimal recommendation performance.} We further empirically validate these observations. Drawing on our investigation, we contend that the aggregation of neighbor nodes facilitates the representational enhancement attributed to GCNs. However, this aggregation process inherently introduces challenges \textbf{C1} and \textbf{C2} during model training. Consequently, we critically point out that only adopting GCNs during the testing phase can enjoy the representational enhancement capabilities of GCNs and effectively circumvent the associated training challenges.

Based on the above findings, we propose an efficient MMRec framework, FastMMRec, which deploys graph convolutions exclusively during the testing phase, bypassing their use in training. FastMMRec can effectively address the scalability problem caused by deploying GCNs during the training phase. Specifically, adopting GCNs exclusively during the testing phase not only prevents constructing useless and even harmful positive and negative pairs and prevents isolation between modalities, but it also retains the representation enhancement capabilities of GCNs through the aggregation of neighboring nodes. To achieve satisfactory performance, we adopt a tailored item-item graph enhancement during the training phase and provide a theoretical analysis to verify that adopting item-item graph enhancement will not lead to the same challenges as the GCNs. We detail the training phase and testing phase implementation of our FastMMRec in Section~\ref{sec: Method}. 

\section{Preliminary}
We conceptualize the user-item interaction graph as $\mathcal{G} = (\mathcal{U}, \mathcal{I}, \mathcal{E})$, where $\mathcal{U}$ and $\mathcal{I}$ denote the collections of users and items, respectively, and $\mathcal{E}$ represents the set of interactions. An edge $(u, i) \in \mathcal{E}$ indicates a user $u$ has interacted with an item $i$. The number of edges is denoted by $|\mathcal{E}|$. To enrich the user-item interaction graph $\mathcal{G}$ with diverse modalities, we introduce modality-specific item embedding $i^m$ for each item $i$ belonging to the set of modalities $\mathcal{M}$. For user and item embedding, $\mathbf{E}_{u^{m}} \in \mathbb{R}^{d \times|\mathcal{U}|}$ represents the user’s randomly initialized embedding, and $\mathbf{E}_{i^m} \in \mathbb{R}^{d \times|\mathcal{I}|}$ represents item initialized embedding with modality $m$, extracted by pre-trained encoders. $d$ signifies the dimensionality of these features. Formally, given an MMRec model denoted as $f(\cdot)$:
\begin{equation}
    s_{u,i} = f(e_u,\{e_{i^m}|m \in \mathcal{M}\} |\Theta),
\end{equation}
where $\Theta$ $\in$ $\mathbb{R}^d$ denotes the model parameters of $f(\cdot)$. Here, $e_u$ and $e_{i^m}$ denote embeddings of user $u$ and item $i$ (with modality $m$), respectively. The predicted score $s_{u, i}$ indicates user $u$'s preference for item $i$, with higher scores reflecting greater interest.
\section{Investigation}
\label{sec: Investigation}
In this section, we first investigate the impact of graph convolution during the training phase, observing that it spreads the optimization of each node in the loss function to its neighboring nodes.
We then identify the first challenge of GCNs that has been overlooked in prior MMRec work: \textbf{C1. GCNs inevitably create unhelpful or even harmful positive and negative pairs during model optimization.} Additionally, we empirically reveal a second challenge in MMRec scenarios: \textbf{C2. GCNs isolate different modalities, resulting in sub-optimal recommendation performance.}
\subsection{What do GCNs actually do during Training?}
Existing studies \cite{wei2019mmgcn,wang2021dualgnn,zhang2021mining,zhou2023tale,zhou2023enhancing} in the MMRec use LightGCN, a lightweight GCN that removes the activation functions and feature transformations of vanilla GCN for each modality $m$. Formally:
\begin{equation}
\resizebox{0.92\hsize}{!}{$\begin{aligned}
e_{u^m}^{(l)}=\sum_{\tilde{i} \in N(u)} \frac{e_{\tilde{i}^{m}}^{(l-1)}}{\sqrt{|N(u)||N(\tilde{i})|}},  \quad e_{i^m}^{(l)}=\sum_{\tilde{u} \in N(i)} \frac{e_{\tilde{u}^{m}}^{(l-1)}}{\sqrt{|N(i)||N(\tilde{u})|}},
\end{aligned}$}
\end{equation}
where $N(\cdot)$ refers to the set of items or users that interact with user $u$ and item $i$, $l$ is the layer number. For the basic Matrix Factorization (MF) model, the similarity $s_{u,i}^m$ for modality $m$ between any user $u$ and item $i$ can be defined as:
\begin{equation}
    s_{u,i}^m = {e_{u^m}}^{\top}  e_{i^m}.
\end{equation}

For a one-layer LightGCN, we unfold the calculation of the similarity $s_{u,i}^m$ for modality $m$ between any user $u$ and item $i$ as follows:
\begin{equation}
\label{eq: GCN1}
\resizebox{1\hsize}{!}{$\begin{aligned}
s_{u,i}^m & =  (e_{u^m}+\sum_{\tilde{i} \in N(u)} \frac{e_{\tilde{i}^{m}}}{\sqrt{|N(u)||N(\tilde{i})|}})^{\top}  (e_{i^m}+\sum_{\tilde{u} \in N(i)} \frac{e_{\tilde{u}^{m}}}{\sqrt{|N(i)||N(\tilde{u})|}}) \\ &
= \underbrace{{e_{u^m}}^{\top}  e_{i^m}}_{\text{Node with Node}}+\underbrace{\sum_{\tilde{u} \in N(i)} \frac{{e_{u^m}}^{\top}  e_{\tilde{u}^{m}}}{\sqrt{|N(i)||N(\tilde{u})|}} + \sum_{\tilde{i} \in N(u)} \frac{{e_{i^m}}^{\top}  e_{\tilde{i}^{m}}}{\sqrt{|N(u)||N(\tilde{i})|}}}_{\text{Node with Neighbors}} \\ &
+ \underbrace{\sum_{\tilde{i} \in N(u)} \sum_{\tilde{u} \in N(i)} \frac{{e_{\tilde{i}^{m}}}^{\top} 
  e_{\tilde{u}^{m}}}{\sqrt{|N(u)||N(\tilde{i})||N(i)||N(\tilde{u})|}}}_{\text{Neighbors with Neighbors}}, 
\end{aligned}$}
\end{equation}
where the final score $s_{u,i}$ is aggregated by $s_{u,i} = \operatorname{Aggr}(s_{u,i}^m | m \in \mathcal{M})$. We divide this function into three parts: \textbf{Node with Node}, \textbf{Node with Neighbors}, and \textbf{Neighbors with Neighbors}. The first part, 'Node with Node' corresponds to the basic interaction mechanism in matrix factorization (MF)-based models. The other two parts, 'Node with Neighbors' and 'Neighbors with Neighbors' reflect the effects of GCNs.


To further analyze what happened in the model optimization process, we first briefly introduce Bayesian Personalized Ranking (BPR) loss \cite{rendle2012bpr}. Essentially, BPR aims to widen the predicted preference margin between positive and negative items for each triplet $(u, p, n) \in \mathcal{D}$, where $\mathcal{D}$ denotes the training set. The positive item $p$ refers to the one with which the user $u$ has interacted, while the negative item $n$ has been randomly chosen from the set of items that the user $u$ has not interacted with. Formally:
\begin{equation}
    \mathcal{L}_{bpr} = \sum_{(u, p, n) \in \mathcal{D}} - \log(\sigma(s_{u,p} - s_{u,n})),
\end{equation}
where $s_{u,p}$ and $s_{u,n}$ are the ratings of user $u$ to the positive item $p$ and negative item $n$. $\sigma(\cdot)$ is the Sigmoid function. For the MF-based models, the BPR loss function directly pulls close the positive pairs, while pushing away the negative pairs, formally:
\begin{equation}
\label{eq: MF2}
    \mathcal{L}_{bpr_{mf}} = \sum_{(u, p, n) \in \mathcal{D}} - \operatorname{Aggr}({e_{u^m}}^{\top}  e_{p^m} - {e_{u^m}}^{\top}  e_{n^m}),
\end{equation}
where we simplify the Sigmoid function and log calculator. For GCN-based models, beyond merely focusing on node pairs, the BPR loss function additionally brings each node and its neighbors closer to the neighbors of its positive node while pushing each node and its neighbors away from the neighbors of its negative node.

We mathematically divided the loss function as:

\begin{equation}
\label{eq: GCN2}
\resizebox{0.92\hsize}{!}{$\begin{aligned}
    \mathcal{L}_{bpr_{gcn}} = & \sum_{(u, p, n) \in \mathcal{D}} - \operatorname{Aggr}(\underbrace{{e_{u^m}}^{\top}  e_{p^m} - {e_{u^m}}^{\top}  e_{n^m}}_{\text{Node $u$ with Node $i$}}
    \\ + & \underbrace{\sum_{\tilde{u}_{p} \in N(p)} \frac{{e_{u^m}}^{\top}  e_{\tilde{u}^{m}_p}}{\sqrt{|N(p)||N(\tilde{u}_p)|}} - \sum_{\tilde{u}_n \in N(n)} \frac{{e_{u^m}}^{\top}  e_{\tilde{u}^{m}_n}}{\sqrt{|N(n)||N(\tilde{u}_n)|}}}_{\text{Node $u$ with Neighbors of $i$}} \\ + &
    \underbrace{\sum_{\tilde{i} \in N(u)} \frac{{e_{p^m}}^{\top}  e_{\tilde{i}^{m}}}{\sqrt{|N(u)||N(\tilde{i})|}} -  \sum_{\tilde{i} \in N(u)} \frac{{e_{n^m}}^{\top}  e_{\tilde{i}^{m}}}{\sqrt{|N(u)||N(\tilde{i})|}}}_{\text{Node $i$ with Neighbors of $u$}}
    \\ + & \underbrace{\sum_{\tilde{i} \in N(u)} \sum_{\tilde{u}_p \in N(p)} \frac{{e_{\tilde{i}^{m}}}^{\top}  e_{\tilde{u}^{m}_p}}{\sqrt{|N(u)||N(\tilde{i})||N(p)||N(\tilde{u}_p)|}}}_{\text{Neighbors of $u$ with Neighbors of $p$}}
    \\ - & \underbrace{\sum_{\tilde{i} \in N(u)} \sum_{\tilde{u}_n \in N(n)} \frac{{e_{\tilde{i}^{m}}}^{\top}  e_{\tilde{u}^{m}_n}}{\sqrt{|N(u)||N(\tilde{i})||N(n)||N(\tilde{u}_n)|}})}_{\text{Neighbors of $u$ with Neighbors of $n$}}.
\end{aligned}$}
\end{equation}

This equation clearly describes how GCN impacts the model optimization. We detailed each part as follows:
\begin{itemize}[leftmargin=*]
    \item \textbf{P1. Node $u$ with Node $i$}: this part directly pulls close the positive pairs while pushing away the negative pairs.
    \item \textbf{P2. Node $u$ with Neighbors of $i$}: this part pulls user closer to other users who interact with its positive item while pushing away user with other users who interact with its negative item.
    \item \textbf{P3. Node $i$ with Neighbors of $u$}: this part pulls positive item closer to other items that interact with its user while pushing away negative item with other items that interact with its user.
    \item \textbf{P4. Neighbors of $u$ with Neighbors of $p$}: this part pulls items that user interacts with closer to users with whom positive item interacts.
    \item \textbf{P5. Neighbors of $u$ with Neighbors of $n$}: this part pushes items the user interacts with away from users with whom negative item interacts.
\end{itemize}

\textbf{P1} reflects the basic assumptions of the BPR loss in the recommendation system. Based on real-world observations, we argue that \textbf{P2-P5} introduce assumptions that are not always beneficial for model optimization. For \textbf{P2}, users $\tilde{u}_n \in N(n)$ who have purchased item $n$, which user $u$ has not bought, do not necessarily have completely different preferences from $u$. For \textbf{P3}, the fact that user $u$ has not purchased item $n$ does not mean that $n$ has attributes completely dissimilar to the items $\tilde{i} \in N(u)$ that $u$ has purchased. For \textbf{P4}, users $\tilde{u}_p \in N(p)$ who have purchased item $p$, which user $u$ also bought, do not necessarily tend to purchase other items $\tilde{i} \in N(u)$ that $u$ has purchased. For \textbf{P5}, users $\tilde{u}_n \in N(n)$ who purchased item $n$, which $u$ has not bought, are not necessarily disinclined to purchase other items $\tilde{i} \in N(u)$ that $u$ has bought.

\begin{figure}[h]
    \centering
    \includegraphics[width=1\linewidth]{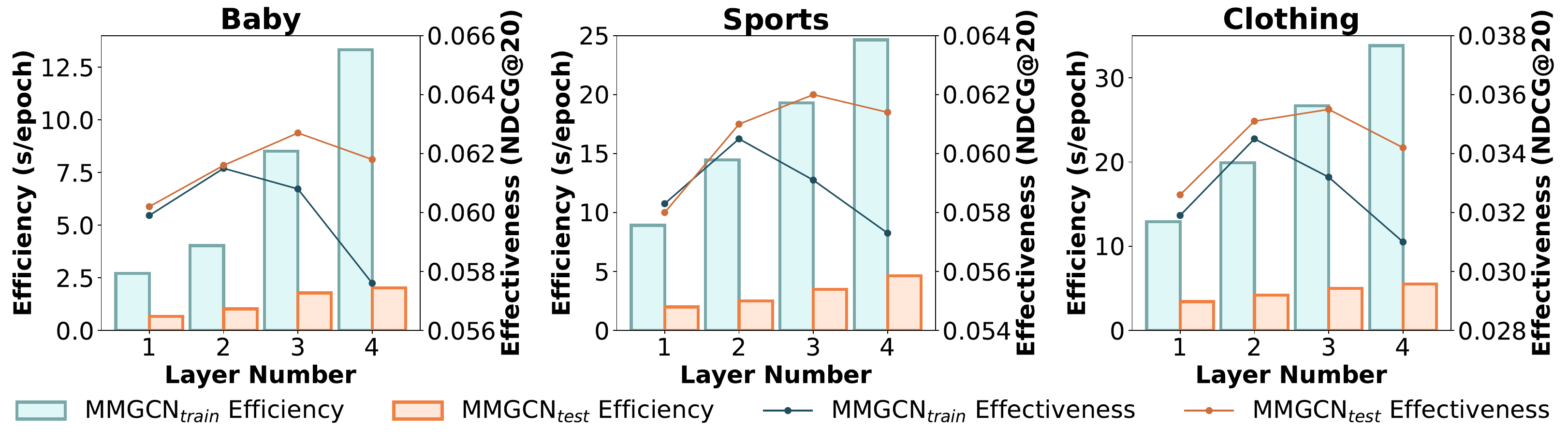}
    \vskip -0.05in
    \caption{Effectiveness and Efficiency study.}   
    \label{fig: MMGCN varaints}
\end{figure}

We further validate our observation through empirical experiments. Specifically, we select MMGCN \cite{wei2019mmgcn}, a widely used GCN-based MMRec model, as the backbone for our study\footnote{To better support our investigation, we provide additional experiments on other advanced MMRec models in Appendix~\ref{appendix: Invest} in supplementary materials.}, and design two variants: MMGCN$_{train}$ and MMGCN$_{test}$. MMGCN$_{train}$ continues to use GCN during the training phase, whereas MMGCN$_{test}$ only adopts GCN during the testing phase, which aims to preserve the powerful representations learned from neighbor aggregation while avoiding the negative influence of bad pairs during the model optimization. We conduct comprehensive experiments\footnote{We use NDCG@20 to evaluate performance and seconds per epoch (s/epoch) to measure efficiency. Details of all metrics and datasets are provided in Section~\ref{sec: Experiment}.} on these two variants, testing different GCN layers across three widely used datasets. As Figure~\ref{fig: MMGCN varaints} shows, MMGCN$_{test}$ outperforms MMGCN$_{train}$ across all datasets and significantly reduces training time. Therefore, we empirically confirm the negative influence of challenge \textbf{C1} and demonstrate that using GCNs exclusively during the testing phase can effectively address this challenge.


In addition to the challenge \textbf{C1} posed by GCNs within each modality, we further examine how GCNs affect the similarity between different modalities. To analyze this, we revisit Equations~\ref{eq: MF2} and \ref{eq: GCN2}. For MF in Equation~\ref{eq: MF2}, the model learns specific weights for each user's modalities by directly optimizing the nodes within each modality. Conversely, for GCN in Equation~\ref{eq: GCN2}, the aggregation of neighbor nodes limits the model's ability to effectively learn specific weights for each modality. Aggregating too many neighbor nodes inevitably reduces the node's unique information, which in turn diminishes personalization \cite{zhou2023layer,zhao2020pairnorm,liu2020towards}. From the overall distribution perspective, this could result in each modality being rigidly tied to its inherent features, consequently leading to modality isolation \cite{yi2024unified,chen2022breaking}.

Therefore, we point out the second challenge faced by GCNs in MMRec: \textbf{C2. GCNs isolate different modalities, resulting in sub-optimal recommendation performance.}

We also validate this observation empirically. Specifically, we measure the similarity between different modalities as follows: 
\begin{equation}
    S=\sum_{o \in (\mathcal{U} \bigcup \mathcal{I})} \frac{S^o}{ |\mathcal{U}| + |\mathcal{I}|}, \quad S^o=\frac{(e_o^v)^{\top} e_o^t}{\|e_o^v\|\|e_o^t\|},
\end{equation}
where $e_o^v$ and $e_o^t$ are visual and textual representations for node $o$. We conduct experiments on three public datasets using the MMGCN variants MMGCN$_{train}$ and MMGCN$_{test}$ to analyze their performance and modality alignment. As shown in Table~\ref{tab: similarity}, the similarity between the visual and textual modalities in MMGCN$_{train}$ is significantly lower than that in MMGCN$_{test}$. This indicates that adopting GCNs during training inevitably isolates different modalities, leading to sub-optimal recommendation performance. This finding provides strong evidence for the negative influence of challenge \textbf{C2}. To further support our findings, we report the similarity $S$ between the visual and textual embeddings for other advanced MMRec models in Appendix~\ref{appendix: Invest} in supplementary materials.

\begin{table}[h]
    \centering
    \small
\caption{Similarity $S$ between visual and textual embeddings.}
\vskip -0.05in
\label{tab: similarity}
    \begin{tabular}{cccc}
    \toprule
         Variants& Baby& Sports& Clothing\\
        \midrule
         MMGCN$_{train}$& 0.2207& 0.2008& 0.2129\\
         MMGCN$_{test}$& 0.3722& 0.3261& 0.3204\\
        \bottomrule
    \end{tabular}
\end{table}

Compared to MMGCN$_{train}$, MMGCN$_{test}$ achieves significant improvements in both efficiency and effectiveness, successfully mitigating these two challenges. However, the state-of-the-art model has a more complex architecture than MMGCN. Consequently, we propose an efficient and high-performing MMRec model, FastMMRec, which adopts GCNs exclusively during the testing phase to achieve superior performance compared to competitive models.

\section{FastMMRec}
\label{sec: Method}
\begin{figure*}
    \centering
    \includegraphics[width=1\linewidth]{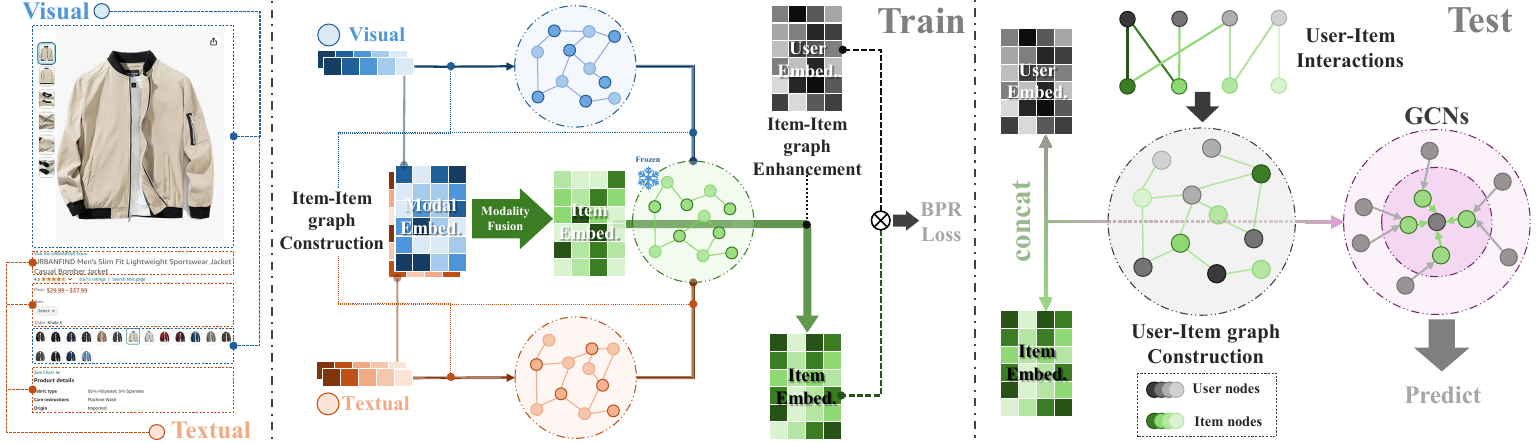}
    \vskip -0.05in
    \caption{The overall framework of the proposed multimodal recommendation model (FastMMRec).}
    \label{fig: architecture}
\end{figure*}

In this section, we detail our FastMMRec for both the training and testing phases. The architecture is depicted in Figure~\ref{fig: architecture}.

\subsection{Training Phase}
To further exploit the rich modality information of items, item-item graphs have been widely used in MMRec \cite{zhou2023tale,zhou2023enhancing,yu2023multi,wei2024llmrec} to aggregate and explore relationships and commonalities among items, achieving satisfactory recommendation performance. We first construct modality-specific item-item graphs using the raw features of each modality (e.g., visual and textual) and then build a unified item-item graph by aggregating all modality-specific graphs. Inspired by previous work \cite{zhou2023tale}, we freeze the similarity graphs during the training phase to reduce computational costs. The pairwise similarity between all items for each modality is calculated as follows:
\begin{equation}
    \mathcal{S}^m_{i,j}=\frac{(e_{i^m})^{\top} e_{j^m}}{\|e_{i^m}\|\|e_{j^m}\|}.
\end{equation}

We retain only the top-$k$ neighbors with the highest similarity scores to capture the most relevant features:
\begin{equation}
\mathcal{\bar{S}}^m_{i,j} = \begin{cases}\mathcal{S}^m_{i,j} & \text { if } \mathcal{S}^m_{i,j} \in \text { top-} k(\mathcal{S}^m_{i,p}|p \in \mathcal{I}) \\ 0 & \text { otherwise }\end{cases},
\end{equation}
where $\mathcal{\bar{S}}^m_{i,j}$ denotes the edge weight between item $i$ and item $j$ within modality $m$. $\mathcal{S}^m_{i,p}|p \in \mathcal{I}$ represents the neighbor scores for the item $i$. To mitigate the issues of gradient explosion or vanishing, we normalize the similarity adjacency matrices as follows:
\begin{equation}
\mathcal{\tilde{S}}^m=(\mathcal{D}^m)^{-\frac{1}{2}} \mathcal{\bar{S}}^m(\mathcal{D}^m)^{-\frac{1}{2}},
\end{equation}
where $\mathcal{D}^m$ is the diagonal degree matrix of $\mathcal{\bar{S}}^m$. Then, we further build a unified item-item graph by aggregating all modality-specific item-item graphs:
\begin{equation}
   \mathcal{\tilde{S}} =  \sum_{m \in \mathcal{M}} \alpha_m \mathcal{\tilde{S}}^m,
\end{equation}
where $\alpha_m$ is a trainable weighted parameter. A unified item-item graph for all modalities can better extract latent relationships across different modalities. Then, we attentively fuse representations of all modalities of users and items, respectively:
\begin{equation}
   \mathbf{E}_{u} = \operatorname{Con}(\alpha_m\mathbf{E}_{u^m}|m \in \mathcal{M}), \quad \mathbf{E}_{i} = \operatorname{Con}(\alpha_m\mathbf{E}_{i^m}|m \in \mathcal{M}),
\end{equation}
where $\operatorname{Con}(\cdot)$ denotes concatenation operation. Then, we aggregate multi-hop neighbors to enhance item representations.
\begin{equation}
   \mathbf{E}_{i} = \mathbf{E}_{i} + \mathbf{E}_{i}(\mathcal{\tilde{S}})^{L_i},
\end{equation}
where $L_i$ is the number of aggregation hop. To preserve the personalization of each item's representation, we add the enhanced representation to the original representation. This strategy ensures that the personalization of each item is maintained while enhanced by aggregating neighbors' representations. 

For model optimization, we compute the inner product of user and item representations to calculate predicted scores and adopt the BPR loss function:
\begin{equation}
    \mathcal{L}_{bpr} = \sum_{(u, p, n) \in \mathcal{D}} - \log(\sigma({e_{u}}^{\top}e_{p} - {e_{u}}^{\top}e_{n})) + \lambda \|\Theta\|_2^2,
\end{equation}
where $\sigma$ is the Sigmoid function. $\lambda$ is a balancing hyper-parameter for regularization terms and $\Theta$ denotes model parameters.

\noindent \textbf{\textit{Analysis: Is item-item graph inevitably build useless and even harmful positive and negative pairs during optimization?}}

We unroll the calculation of the similarity $s_{u,i}$ between any user $u$ and item $i$ with the one-hop item-item graph enhancement:
\begin{equation}
s_{u,i}  = (e_{u})^{\top} (e_{i}+\sum_{\hat{i}| \mathcal{\bar{S}}_{i,\hat{i}} \neq 0} \frac{e_{\hat{i}}}{k})  = \underbrace{{e_{u}}^{\top} e_{i}}_{\text{Node with Node}} +  \underbrace{\sum_{\hat{i}| \mathcal{\bar{S}}_{i,\hat{i}} \neq 0}\frac{{e_{u}}^{\top}e_{\hat{i}}}{k}.}_{\text{Node $u$ with Neighbors of $i$}}
\end{equation}

We divided this function into two parts: \textbf{interactions between Node $u$ and Node $i$} and \textbf{interactions between Node $u$ and the neighbors of Node $i$}. For item-item graph enhancement, the neighbors of item $i$ are other items rather than users, which distinguishes this approach from traditional GCNs. Next, we mathematically analyze the impact of item-item graph enhancement on BPR loss: 

\begin{equation}
\begin{aligned}
    \mathcal{L}_{bpr_{FastMMRec}} = & \sum_{(u, p, n) \in \mathcal{D}} - (\underbrace{{e_{u}}^{\top}  e_{p} - {e_{u}}^{\top}  e_{n}}_{\text{Node $u$ with Node $i$}}
    \\ + & \underbrace{\sum_{\hat{p}| \mathcal{\bar{S}}_{p,\hat{p}}} \frac{{e_{u}}^{\top}  e_{\hat{p}}}{k} - \sum_{\hat{n}| \mathcal{\bar{S}}_{n,\hat{n}}} \frac{{e_{u}}^{\top}  e_{\hat{n}}}{k}}_{\text{Node $u$ with Neighbors of $i$}}).
\end{aligned}
\end{equation}

This equation clearly describes how GCN impacts model optimization. We detailed each part as follows:
\begin{itemize}[leftmargin=*]
    \item \textbf{P1. Node $u$ with Node $i$}: this part directly pulls close the positive pairs while pushing away the negative pairs.
    \item \textbf{P2. Node $u$ with Neighbors of $i$}: this part pulls the item closer to other items that are semantically similar to positive items, while pushing the item away from other items that are semantically similar to negative items.
\end{itemize}

Since the neighbors in an item-item graph are semantically related and consist solely of items, they do not encounter the inherent semantic discrepancies between items and users that are observed in GCNs. As a result, using an item-item graph during training does not create irrelevant or harmful positive and negative pairs during model optimization. Furthermore, it leverages the rich multimodal information of items to enhance the model's robustness. We empirically validate this observation in Section~\ref{sec: RQ2}. 

\subsection{Testing Phase}
Based on the analysis in Section~\ref{sec: Investigation}, we only adopt GCNs during the testing phase to address the challenges associated with employing GCNs during the training phase and to enhance model efficiency. During the testing phase, we utilize GCNs, and the predicted score $s_{u,i}$ between user $u$ and item $i$ is calculated as:
\begin{equation}
\resizebox{0.9\hsize}{!}{$\begin{aligned}
 e_{u}^{(l)}=\sum_{\tilde{i} \in N(u)} \frac{e_{\tilde{i}}^{(l-1)}}{\sqrt{|N(u)||N(\tilde{i})|}}, \quad e_{i}^{(l)}=\sum_{\tilde{u} \in N(i)} \frac{e_{\tilde{u}}^{(l-1)}}{\sqrt{|N(i)||N(\tilde{u})|}},
\end{aligned}$}
\end{equation}
\begin{equation}
\tilde{e}_u = \sum_{l=1}^L e_u^{(l)}, \quad \tilde{e}_i = \sum_{l=1}^L e_i^{(l)}, \quad s_{u,i} = \tilde{e}_{u}^{\top}  \tilde{e}_{i}.
\end{equation}

Adopting GCNs in the testing leverages neighbor representations while avoiding the challenges of using GCNs during training.

\begin{table}[!ht]
    \centering
\caption{Statistics of three experimented datasets with multimodal item Visual(V) and Textual(T) contents.}
\small
\setlength{\tabcolsep}{1.5mm}
 \vskip -0.05in
\label{tab: dataset_statistics}
    \begin{tabular}{cccccccccc}
    \hline
         Dataset&& \multicolumn{2}{c}{Baby} && \multicolumn{2}{c}{Sports} && \multicolumn{2}{c}{Clothing}\\
         \hline \hline`
         Modality && V & T && V & T && V & T\\ 
         Embed Dim && 4096 & 384 && 4096 & 384 && 4096 & 384\\ \cline{1-1} \cline{3-4} \cline{6-7} \cline{9-10}
         User && \multicolumn{2}{c}{19445} && \multicolumn{2}{c}{35598} && \multicolumn{2}{c}{39387} \\
         Item && \multicolumn{2}{c}{7050} && \multicolumn{2}{c}{18357} && \multicolumn{2}{c}{23033} \\
         Interaction && \multicolumn{2}{c}{160792} && \multicolumn{2}{c}{296337} && \multicolumn{2}{c}{278677} \\ \hline
         Sparsity && \multicolumn{2}{c}{99.88\%} && \multicolumn{2}{c}{99.95\%} && \multicolumn{2}{c}{99.97\%} \\
         \hline
    \end{tabular}
\end{table}
\section{Evaluation}
\label{sec: Experiment}

\subsection{Experiment Settings}

\subsubsection{Datasets}
The experiments are conducted on three real-world datasets from the Amazon \cite{mcauley2015image}: Baby, Sports, and Clothing, each encompassing visual and textual modalities for every item. Consistent with most previous studies \cite{zhou2023tale,zhou2023bootstrap,xu2024aligngroup}, we apply the 5-core setting to filter users and items within each dataset. We follow the same setting mentioned in \cite{zhou2023mmrecsm}, which extracts 4096-dimensional visual features and 384-dimensional textual features via pre-trained encoders. Table~\ref{tab: dataset_statistics} presents the statistics of these datasets. For each dataset, we randomly split the historical interactions using an 8:1:1 ratio for training, validation, and testing.

\subsubsection{Baselines}
To verify the effectiveness of our proposed FastMMRec, we compare FastMMRec with a variety of baselines, including conventional recommendation models (\textbf{MF-BPR} \cite{rendle2012bpr}, \textbf{LightGCN} \cite{he2020lightgcn}, \textbf{SimGCL} \cite{yu2022graph}, and \textbf{LayerGCN} \cite{zhou2023layer}) and multimodal recommendation models (\textbf{VBPR} \cite{he2016vbpr}, \textbf{MMGCN} \cite{wei2019mmgcn}, \textbf{DualGNN} \cite{wang2021dualgnn}, \textbf{LATTICE} \cite{zhang2021mining}, \textbf{FREEDOM} \cite{zhou2023tale}, \textbf{SLMRec} \cite{tao2022self}, \textbf{BM3} \cite{zhou2023bootstrap}, \textbf{MMSSL} \cite{wei2023multi}, \textbf{LGMRec} \cite{guo2024lgmrec}, and \textbf{DiffMM} \cite{jiang2024diffmm}). Details of baselines are presented in Appendix~\ref{appendix: baseline} in supplementary materials.

\subsubsection{Metrics}
To evaluate the top-K recommendation task performance fairly, we adopt two widely-used metrics: Recall and NDCG. We report the average metrics of all users in the test dataset under Recall@10 (R@10), Recall@20 (R@20), NDCG@10 (N@10), and NDCG@20 (N@20).

\subsubsection{Implementation Details}
To ensure a fair comparison, we implement our FastMMRec and all the baselines using the MMRec library \cite{zhou2023mmrecsm}. Specifically, all models are implemented in PyTorch, using the Adam optimizer \cite{kingma2014adam} and Xavier initialization \cite{glorot2010understanding} with default parameters. We perform a complete grid search for all baselines to determine their optimal settings as described in their published papers. As for hyper-parameter settings on our FastMMRec, we perform a grid search on the item-item graph hop number $L_i$ in \{1, 2, 3\},  the $k$ of top-$k$ item-item graph in \{5, 10, 15, 20\}, and the layer number $L$ of GCN in \{1, 2, 3, 4\}. We empirically fix the learning rate with $1e^{-4}$ and regularization weight $\lambda$ with $1e^{-3}$. To avoid the over-fitting problem, we set 20 as the early stopping epoch number. Following previous studies \cite{zhou2023mmrecsm,xu2025mentor}, we utilize Recall@20 on the validation dataset as a metric to update the best record. Note that all models are evaluated on an RTX 3090 with 24GB memory.

\begin{table*}[!ht]
\caption{Performance comparison of baselines on different datasets in terms of \emph{Recall}@K and \emph{NDCG}@K.}
 \vskip -0.05in
 \small
\centering
\label{tab: RQ1}
\setlength{\tabcolsep}{1.55mm}
\resizebox{\linewidth}{!}{
    \begin{tabular}{ccccclcccclcccc}
    \hline
         \multirow{2}{*}{Baseline} &  \multicolumn{4}{c}{Baby}&& \multicolumn{4}{c}{Sports}&&  \multicolumn{4}{c}{Clothing}\\\cline{2-5} \cline{7-10} \cline{12-15} 
         & R@10& R@20& N@10& N@20&& R@10& R@20& N@10& N@20&& R@10& R@20& N@10& N@20\\\hline
         MF-BPR & 0.0357& 0.0575& 0.0192& 0.0249&& 0.0432& 0.0653& 0.0241& 0.0298&& 0.0187& 0.0279& 0.0103& 0.0126\\
         LightGCN & 0.0479& 0.0754& 0.0257& 0.0328&& 0.0569& 0.0864& 0.0311& 0.0387&& 0.0340& 0.0526& 0.0188& 0.0236\\
         SimGCL & 0.0513& 0.0804& 0.0273& 0.0350&& 0.0601& 0.0919& 0.0327& 0.0414&& 0.0356& 0.0549& 0.0195& 0.0244\\
         LayerGCN & 0.0529& 0.0820& 0.0281& 0.0355&& 0.0594& 0.0916& 0.0323& 0.0406&& 0.0371& 0.0566& 0.0200& 0.0247\\
         \hline
         VBPR & 0.0423& 0.0663& 0.0223& 0.0284&& 0.0558& 0.0856& 0.0307& 0.0384&& 0.0281& 0.0415& 0.0158& 0.0192\\
         MMGCN & 0.0378& 0.0615& 0.0200& 0.0261&& 0.0370& 0.0605& 0.0193& 0.0254&& 0.0218& 0.0345& 0.0110& 0.0142\\
         DualGNN & 0.0448& 0.0716& 0.0240& 0.0309&& 0.0568& 0.0859& 0.0310& 0.0385&& 0.0454& 0.0683& 0.0241& 0.0299\\
         LATTICE & 0.0547& 0.0850& 0.0292& 0.0370&& 0.0620& 0.0953& 0.0335& 0.0421&& 0.0492& 0.0733& 0.0268& 0.0330\\
         FREEDOM & 0.0627& \underline{0.0992}& 0.0330& 0.0424&& 0.0717& \underline{0.1089}& 0.0385& \underline{0.0481}&& \underline{0.0628}& \underline{0.0941}& \underline{0.0341}& \underline{0.0420}\\
         SLMRec & 0.0529& 0.0775& 0.0290& 0.0353&& 0.0663& 0.0990& 0.0365& 0.0450&& 0.0452&  0.0675& 0.0247& 0.0303\\
         BM3 & 0.0564& 0.0883& 0.0301& 0.0383&& 0.0656& 0.0980& 0.0355& 0.0438&& 0.0422& 0.0621& 0.0231& 0.0281\\
         MMSSL & 0.0613& 0.0971& 0.0326& 0.0420&& 0.0673& 0.1013& 0.0380& 0.0474&& 0.0531& 0.0797& 0.0291& 0.0359\\
         LGMRec & \underline{0.0639}& 0.0989& \underline{0.0337}& \underline{0.0430}&& \underline{0.0719}& 0.1068& \underline{0.0387}& 0.0477&& 0.0555& 0.0828& 0.0302& 0.0371\\
         DiffMM & 0.0623& 0.0975& 0.0328& 0.0411&& 0.0671& 0.1017& 0.0377& 0.0458&& 0.0522& 0.0791& 0.0288& 0.0354\\
         \hline
         FastMMRec & \textbf{0.0667}& \textbf{0.1034}& \textbf{0.0357}& \textbf{0.0453}&& \textbf{0.0768}& \textbf{0.1151}& \textbf{0.0415}& \textbf{0.0517}&& \textbf{0.0674}& \textbf{0.0992}& \textbf{0.0366}& \textbf{0.0447}\\
         $p$-value & 2.21$e^{-4}$& 8.33$e^{-5}$& 1.01$e^{-4}$& 1.85$e^{-4}$&& 4.60$e^{-4}$& 3.11$e^{-4}$& 5.28$e^{-4}$& 5.51$e^{-4}$&& 4.94$e^{-4}$& 2.89$e^{-4}$& 5.42$e^{-4}$& 4.67$e^{-4}$\\
         Improv.& 4.38\%& 4.23\%& 5.93\%& 5.35\%&& 6.82\%& 5.89\%& 7.24\%& 7.48\%&& 7.32\%& 5.42\%& 7.33\%& 6.43\%\\\hline
    \end{tabular}
    }
\end{table*}

\subsection{Performance Comparison}
Table~\ref{tab: RQ1} presents the evaluation results of the performance comparison. In this table, we highlight the optimal results in bold and underline the sub-optimal results for easy identification. We have the following key observations:

\begin{itemize}[leftmargin=*]
    \item \textbf{Performance superiority of our FastMMRec.} Our FastMMRec consistently outperforms all baselines across diverse datasets. This advantage is attributable to the fact that we adopt GCNs only during the testing phase. This approach avoids creating irrelevant or harmful positive and negative pairs during model optimization and prevents isolation between modalities.
    \item \textbf{Effectiveness of item-item graph enhancement.} Utilizing item-item graph enhancement significantly improves the performance of recommender systems. Models such as LATTICE, FREEDOM, MMSSL, LGMRec, and FastMMRec benefit significantly from this approach. This improvement is due to item-item graphs, which enhance item representations by aggregating semantically related neighbors and extracting rich multimodal information.
    \item \textbf{Importance of multimodal information.} Most multimodal recommendation models outperform conventional ones, highlighting the importance of incorporating multimodal information to learn user preferences and item properties.
\end{itemize}

\begin{table}[!ht]
\caption{Ablation study on key components of FastMMRec in terms of \emph{Recall}@20 and \emph{NDCG}@20.}
 \vskip -0.05in
 \small
\centering
\label{tab: RQ2}
\resizebox{\linewidth}{!}{
    \begin{tabular}{c|cc|cc|cc}
    \hline
         Dataset &  \multicolumn{2}{c|}{Baby}& \multicolumn{2}{c|}{Sports}&  \multicolumn{2}{c}{Clothing} \\\hline
         Variants & Recall& NDCG& Recall& NDCG& Recall& NDCG \\\hline\hline
         \emph{w/o-item} & 0.0947& 0.0402& 0.1035& 0.0461& 0.0899& 0.0404\\
         \emph{test-item} & 0.1013& 0.0434& 0.1119& 0.0502& 0.0963& 0.0431\\\hline
         FastMMRec & \textbf{0.1034}& \textbf{0.0453}& \textbf{0.1151}& \textbf{0.0517}& \textbf{0.0992}& \textbf{0.0447}\\\hline
    \end{tabular}
    }
\end{table}

\subsection{Ablation Study}
\label{sec: RQ2}
To validate the effectiveness of FastMMRec, we conduct experiments to justify the importance of key components. We design the following variants: 1) \emph{w/o-item}, which removes the item-item graph entirely, and 2) \emph{test-item}, which applies the item-item graph enhancement only during the testing phase instead of the training phase. The results in Table~\ref{tab: RQ2} provides following key conclusions:

\begin{itemize}[leftmargin=*]
    \item The performance of \emph{w/o-item} drops significantly compared to FastMMRec, demonstrating the effectiveness of the item-item graph enhancement component.
    \item Comparing FastMMRec with \emph{test-item}, we observe a clear performance advantage, reflecting the benefits of using item-item graph enhancement during model optimization.
\end{itemize} 


\begin{figure}[h]
    \centering
    \includegraphics[width=1\linewidth]{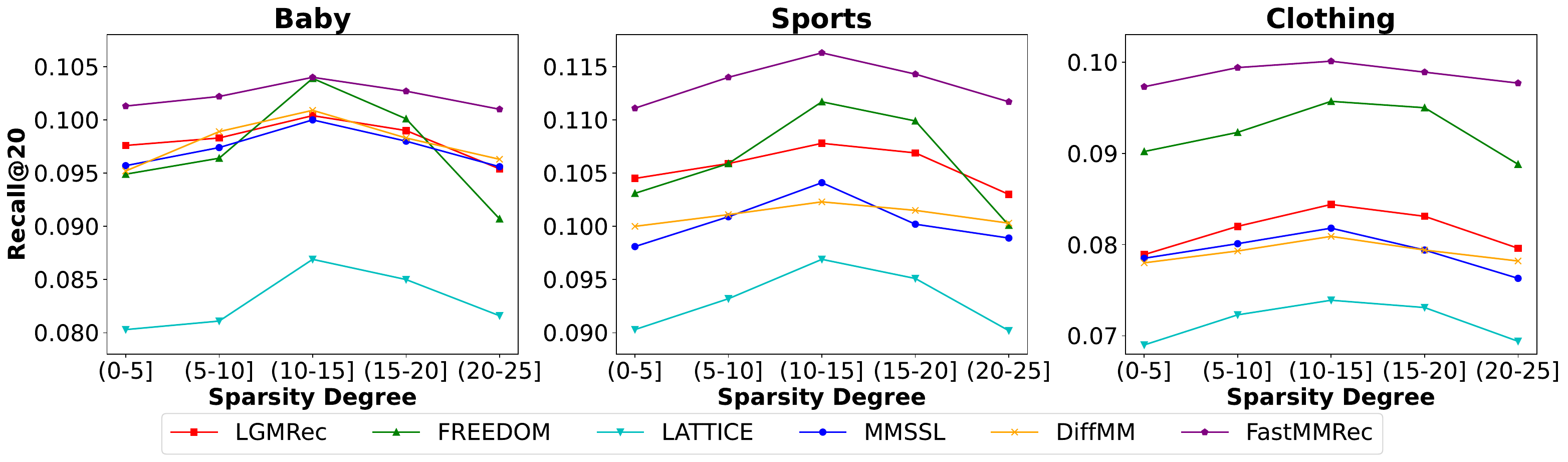}
    \vskip -0.05in
    \caption{Sparsity degree analysis on three datasets.}   
    \label{fig: RQ3}
\end{figure}

\subsection{Sparsity Study}
We validate the effectiveness of FastMMRec under different levels of data sparsity. To assess its performance, we conduct experiments on sub-datasets derived from all three datasets, each with varying levels of data sparsity. We compare FastMMRec's performance with five competitive baselines: LATTICE, FREEDOM, MMSSL, LGMRec, and DiffMM. We categorize user groups based on the number of interactions in the training set, such as users with 0–5 interacted items in the first group. Figure~\ref{fig: RQ3} shows that FastMMRec consistently outperforms all baselines across datasets, confirming its robustness under different sparsity levels.

\begin{table*}[!ht]
\caption{Comparison of computational complexity on graph-based multimodal models.}
 \vskip -0.05in
 \small
\centering
\label{tab: RQ4-1}
\resizebox{\linewidth}{!}{
    \begin{tabular}{c|c|c|c|c}
    \hline
        Stage & MMGCN & LATTICE & MMSSL & FastMMRec \\ \hline
        Graph Convolution & $O(2|\mathcal{M}|L|\mathcal{E}|d/B)$ & $O(2L|\mathcal{E}|d/B)$ & $O(2|\mathcal{M}|L|\mathcal{E}|d/B)$ & - \\ \hline
        Feature Mapping & $O(\sum_{m \in \mathcal{M}}|I|(d_m+d) d_h)$ & $O(|\mathcal{I}|^3+\sum_{m \in \mathcal{M}}|\mathcal{I}|^2 d_m+k|\mathcal{I}| \log (|\mathcal{I}|))$ & $O(\sum_{m \in \mathcal{M}}|\mathcal{I}| d_m d)$ & $O(|\mathcal{I}| d^2)$ \\ \hline
        Loss & $O(2dB)$ & $O(2dB)$ & $O((2+|\mathcal{M}||\mathcal{U}||\mathcal{I}|+2|\mathcal{M}|) d B+|\mathcal{M}||\mathcal{U}||\mathcal{I}| d_m B)$ & $O(2dB)$ \\ \hline
    \end{tabular}
    }
    \\
    $d_h$ denotes the dimension of the hidden layer in a two-layer MLP and $k$ is the value of top-$k$ neighbors in the item-item graph.
\end{table*}

\begin{table*}
    \centering
\caption{Comparison of our FastMMRec against state-of-the-art baselines on model efficiency.}
\vskip -0.05in
\label{tab: RQ4-2}
\resizebox{\linewidth}{!}{
    \begin{tabular}{llccccccccccc}\hline
        Dataset& Metrics & VBPR & MMGCN & DualGNN & LATTICE & FREEDOM & SLMRec & BM3 & MMSSL & LGMRec & DiffMM & FastMMRec \\
        \hline
        \multirow{2}{*}{Baby} 
        & Time (s/epoch)& 0.55& 4.09& 5.63& 3.20& 2.57& 2.07& 1.93& 6.31& 4.19& 9.45& 0.61\\
        & Memory (GB)& 1.89& 2.69& 2.05& 4.53& 2.13& 2.08& 2.11& 3.77& 2.41& 4.23& 1.93\\
        \hline
        \multirow{2}{*}{Sports} 
        & Time (s/epoch)& 0.97& 14.93& 11.59& 11.07& 5.65& 5.39& 3.82& 14.67& 8.38& 18.61& 1.01\\
        & Memory (GB)& 2.71& 3.91& 2.81& 19.93& 3.34& 3.04& 3.58& 5.34& 3.67& 5.99& 2.79\\ 
        \hline
        \multirow{2}{*}{Clothing} 
        & Time (s/epoch)& 1.34& 17.48& 14.19& 16.53& 6.29& 6.02& 5.25& 17.04& 9.72& 23.85& 1.39\\
        & Memory (GB)& 3.02& 4.24& 3.02& 28.22& 4.15& 3.40& 4.13& 5.81& 4.81& 6.54& 3.11\\
        \hline
    \end{tabular}
    }
\end{table*}

\subsection{Efficiency Study}
Our FastMMRec achieves surprising efficiency improvements compared to previous studies. We analyze its efficiency by complexity, convergence, and training time. 

\subsubsection{Complexity} Our FastMMRec achieves significant efficiency improvements over previous studies. We analyze the efficiency of our FastMMRec in terms of complexity, convergence, and training time. We analyze the computational complexity of FastMMRec and compare it with three advanced graph-based MMRec models (MMGCN, LATTICE, and MMSSL) in Table~\ref{tab: RQ4-1}. We divide computational complexity into three major components: \textbf{Graph Convolution}, \textbf{Feature Mapping}, and \textbf{Loss}. \textbf{1) Graph Convolution.} MMGCN and MMSSL adopt LightGCN for each modality, with a computational complexity of $O(2|\mathcal{M}|L|\mathcal{E}|d/B)$, where $|\mathcal{M}|$ is the number of modalities, $L$ is the number of layers in LightGCN, $B$ is the batch size, $d$ is the embedding dimension, and $|\mathcal{E}|$ is the number of edges in the graph. LATTICE adopts a single LightGCN for the fused modality, with a computational complexity of $O(2L|\mathcal{E}|d/B)$. For FastMMRec, we only adopt GCNs in the testing phase, eliminating all computational costs for graph convolution. \textbf{2) Feature Mapping.} MMGCN uses a two-layer MLP feature projection for each modality, with a complexity of $O(\sum_{m \in \mathcal{M}}|\mathcal{I}|(d_m + d) d_h)$, where $d_h$ is the hidden dimension and $|\mathcal{I}|$ is the number of items. LATTICE constructs an item-item graph from multimodal features, which involves $O(\sum_{m \in \mathcal{M}}|\mathcal{I}|^2 d_m)$ to build the similarity matrix, $O(|\mathcal{I}|^3)$ to normalize the matrix, and $O(k|\mathcal{I}| \log(|\mathcal{I}|))$ to retrieve the top-$k$ most similar items, where $k$ is the number of neighbors per item. MMSSL freezes the item-item graph during training, with a complexity of $O(|\mathcal{I}|d_m d)$ per modality, resulting in a total complexity of $O(\sum_{m \in \mathcal{M}}|\mathcal{I}|d_m d)$. FastMMRec also freezes the item-item graph during training and uses a single fused item-item graph, resulting in a total complexity of $O(|\mathcal{I}|d^2)$. \textbf{3) Loss.} MMGCN, LATTICE, and FastMMRec use the vanilla BPR loss, with a complexity of $O(2dB)$. MMSSL, in addition to vanilla BPR loss ($O(2dB)$), includes generator loss ($O(|\mathcal{M}||\mathcal{U}||\mathcal{I}|dB)$), discriminator loss ($O(|\mathcal{M}||\mathcal{U}||\mathcal{I}|d_m B)$), and contrastive learning loss ($O(2|\mathcal{M}|dB)$). \textbf{FastMMRec entirely avoids the complex Graph Convolution module, exhibits lower complexity in the Feature Mapping module compared to existing methods, and only incurs the complexity of the vanilla BPR loss in the Loss module.}


\begin{figure}[h]
    \centering
    \includegraphics[width=1\linewidth]{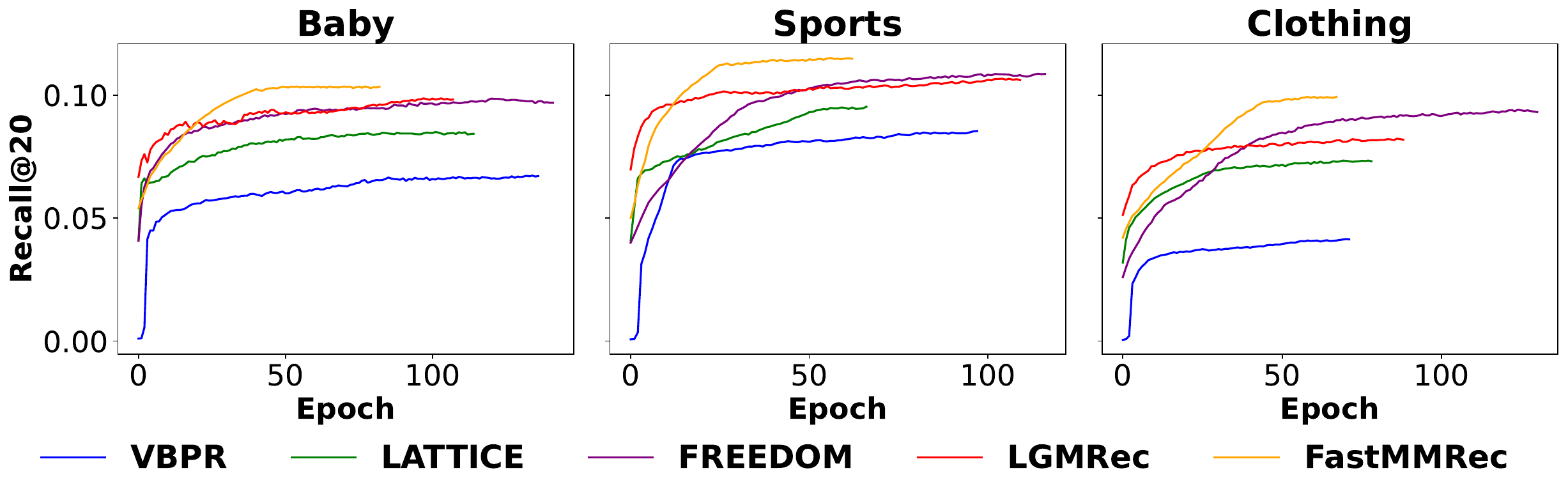}
    \vskip -0.05in
    \caption{Convergence study in terms of \emph{Recall}@20.}
    \label{fig: RQ4}
\end{figure}

\subsubsection{Convergence}
Figure~\ref{fig: RQ4} shows the training curves of our FastMMRec and the compared models (VBPR, LATTICE, FREEDOM, and LGMRec) on all three datasets as the number of iterations and epochs increases. We have the following findings:
\begin{itemize}[leftmargin=*]
\item The faster convergence speed of FastMMRec is clearly evident, highlighting its advantage in training efficiency while maintaining superior recommendation accuracy. This suggests that adopting GCNs only during the testing phase facilitates faster convergence during model training.
\item FastMMRec and VBPR achieve faster convergence speeds than graph-based models (LATTICE and FREEDOM), further confirming that GCNs often construct irrelevant or even harmful positive and negative pairs during model optimization. By adopting GCNs only during the testing phase, FastMMRec addresses this challenge and leverages the representational benefits of GCNs.
\end{itemize}

\subsubsection{Training time}
We report the training time and memory usage of FastMMRec and baselines in Table~\ref{tab: RQ4-2}. We make the following observations:

\begin{itemize}[leftmargin=*]
    \item \textbf{Training time:} FastMMRec demonstrates faster training speeds, while other graph-based models show a rapid increase in training time as dataset size grows. In contrast, FastMMRec scales approximately linearly with dataset size. This efficiency is due to the exclusive adoption of GCNs during the testing phase, effectively addressing the scalability challenges of graph-based models in real-world applications.
    \item \textbf{Memory:} FastMMRec uses less memory than all other graph-based models, attributable to FastMMRec constructing a single item-item graph and freezing it during the training phase.
\end{itemize}

\begin{figure}[h]
    \centering
      \vskip -0.15in
    \subfigure[FREEDOM] {
        \label{fig: RQ5-1}
        \includegraphics[width=0.48\linewidth]{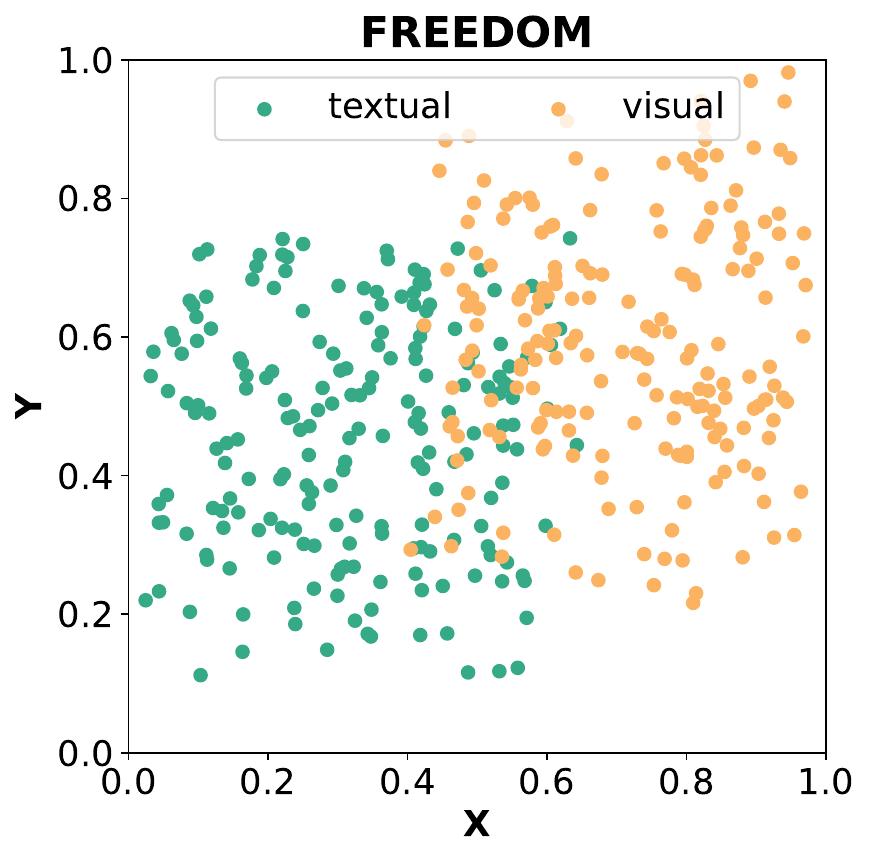}
        }  \hspace{-3.mm}
    \subfigure[FastMMRec] {
        \label{fig: RQ5-2}
        \includegraphics[width=0.48\linewidth]{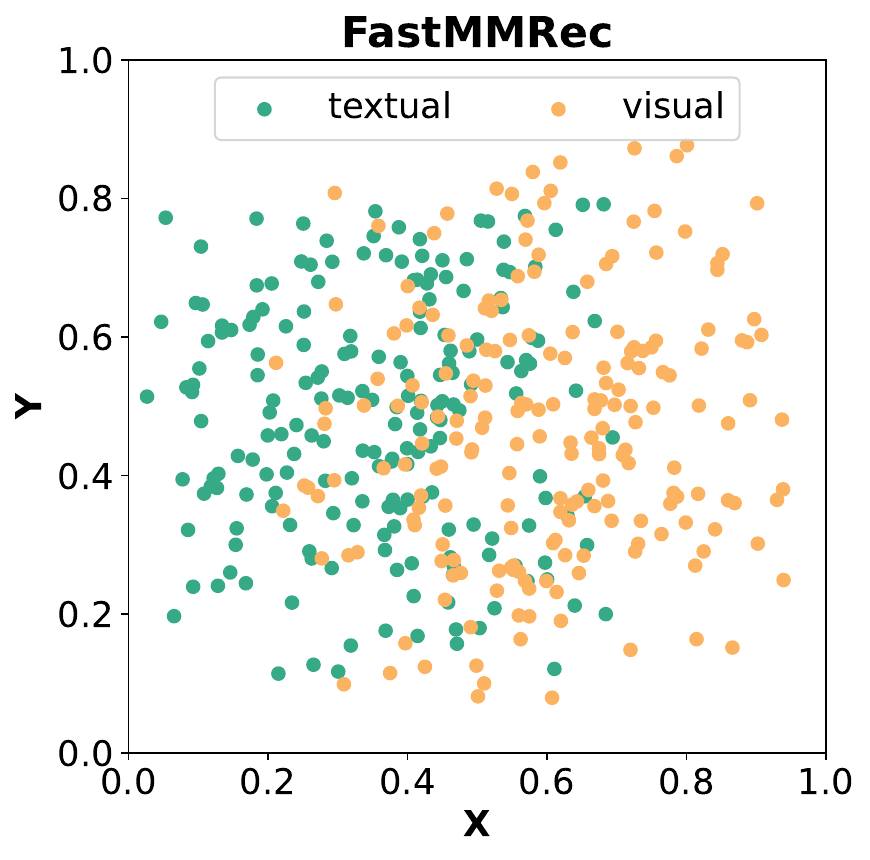}
        }  
    \vskip -0.15in
    \caption{Distribution of visual and textual representations obtained by FREEDOM and FastMMRec on the Baby dataset.}   
    \label{fig: RQ5}
 \vskip -0.15in
\end{figure}

\subsection{Visualization}
To further validate the advantages of FastMMRec in preventing the modality isolation problem, we perform the following analysis: We randomly select 200 items from the Baby dataset and apply the t-SNE \cite{van2008visualizing} to project the item representations of FREEDOM and FastMMRec into a two-dimensional space. Upon analyzing the 2D feature distributions in Figure~\ref{fig: RQ5}, we observe that the visual and textual feature distributions in FastMMRec are more similar to each other compared to FREEDOM. This similarity suggests that FastMMRec effectively mitigates the modality isolation problem.

\begin{figure}[h]
    \centering
    \includegraphics[width=1\linewidth]{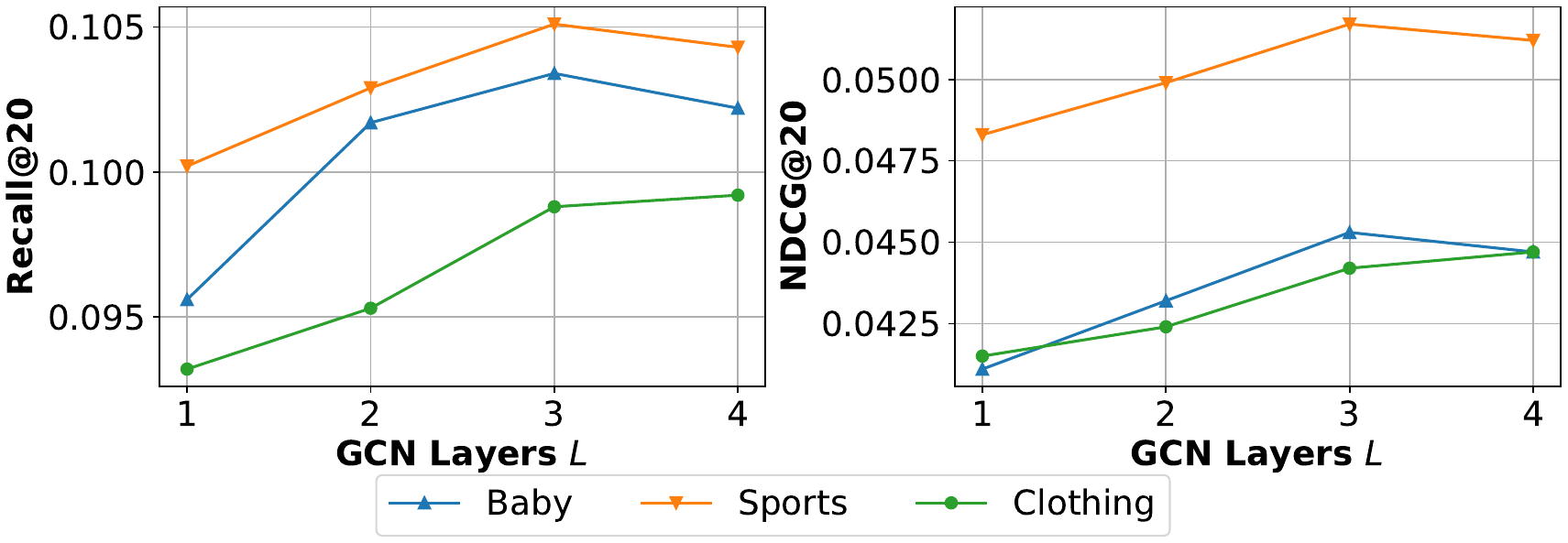}
    \vskip -0.05in
    \caption{Effect of GCN layers $L$.}
    \label{fig: RQ6-1}
\end{figure}

\begin{figure}[h]
    \centering
    \subfigure[Baby] {
        \label{fig: baby}
        \includegraphics[width=0.32\linewidth]{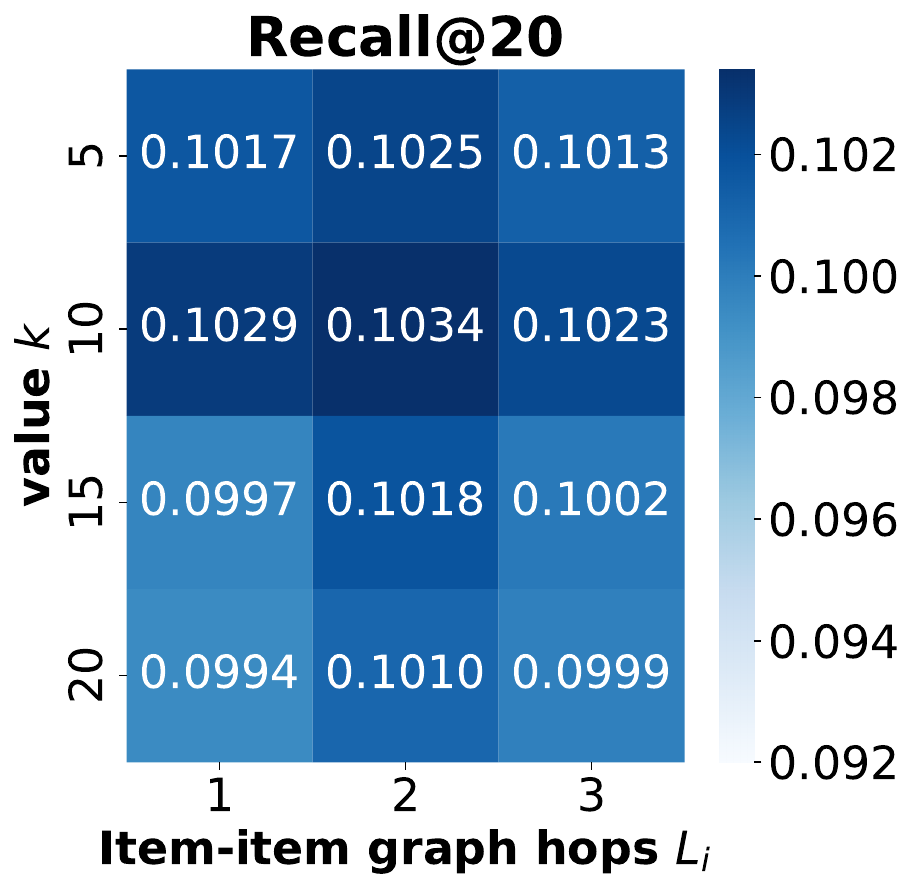}
        }  \hspace{-4.mm}
    \subfigure[Sports] {
        \label{fig: sports}
        \includegraphics[width=0.32\linewidth]{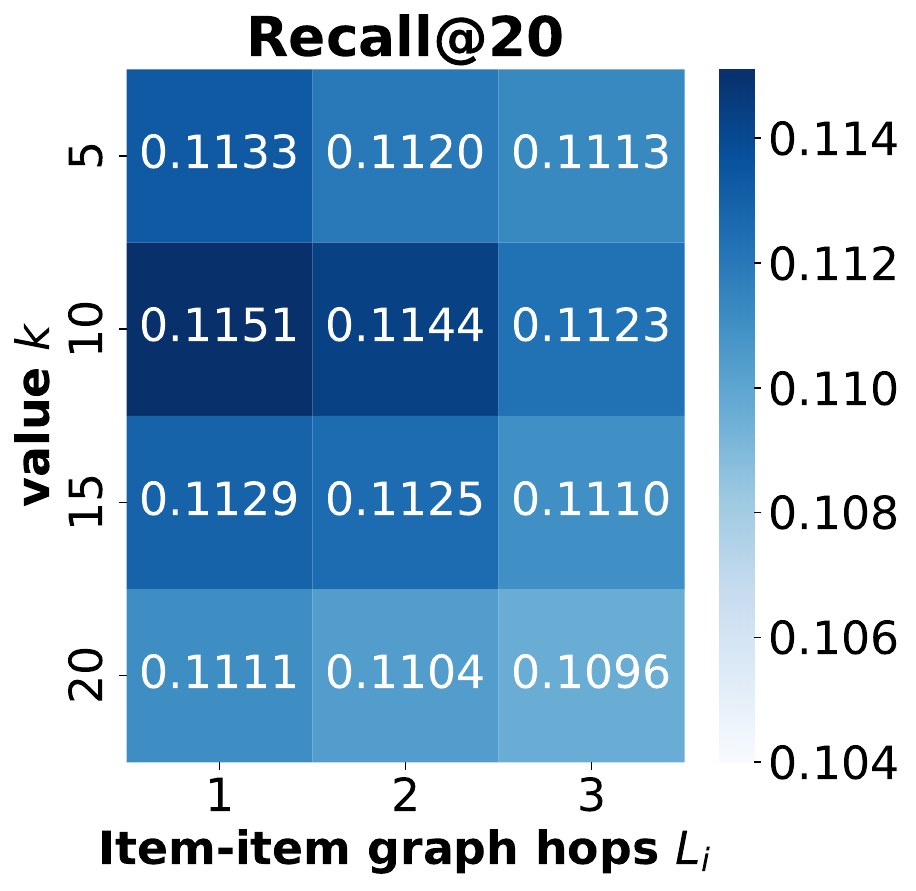}
        }  \hspace{-4.mm}
     \subfigure[Clothing] {
        \label{fig: clothing}
        \includegraphics[width=0.32\linewidth]{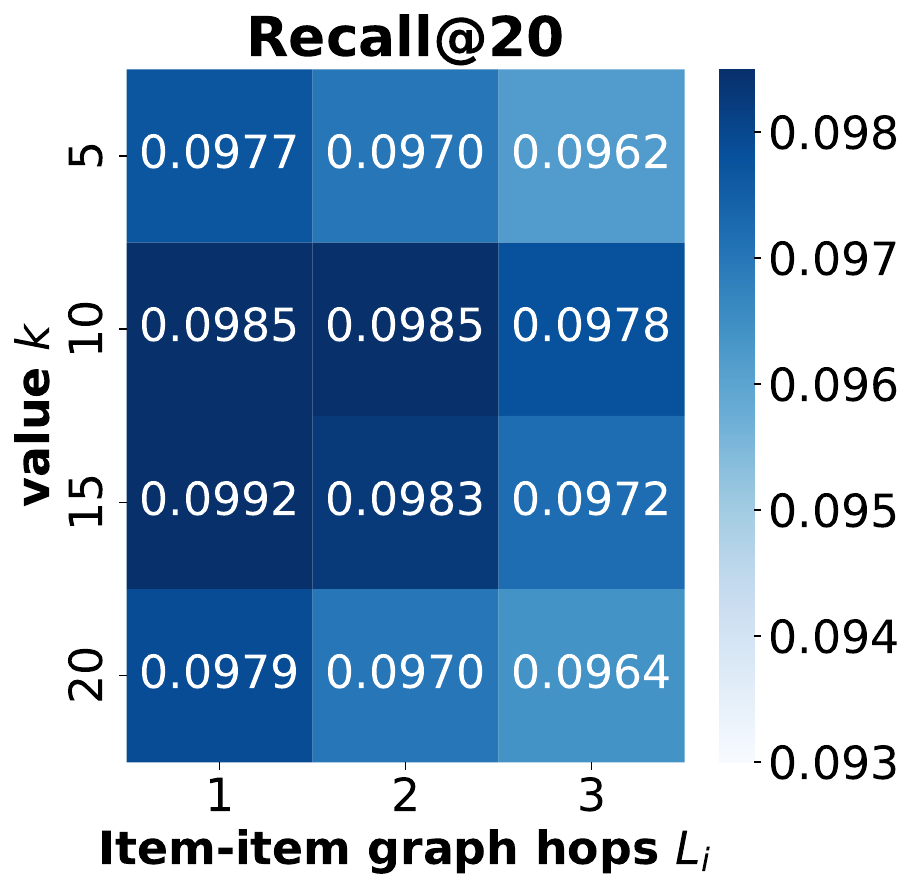}
        } 
    \vskip -0.05in
    \caption{Effect of item-item graph hops $L_i$ and $k$ value.}   
    \label{fig: RQ6-2}
\end{figure}

\subsection{Hyper-parameter Study}
We examine the sensitivity of several important hyper-parameters of FastMMRec across different datasets.

\begin{itemize}[leftmargin=*]
    \item \textbf{GCN layers $L$:} We first investigate the impact of GCN depth by varying the number of message-passing layers $L$ in $\{1,2,3,4\}$. Figure~\ref{fig: RQ6-1} shows FastMMRec achieves its best performance with $L$ = 3 or 4. Note that existing MMRec models suffer from performance deterioration due to the over-smoothing problem when the number of GCN layers reaches 2 or 3. FastMMRec mitigates this problem by adopting GCNs only during the testing phase.
    \item \textbf{Item-item graph hops $L_i$ and $k$ value:} We empirically analyze the impact of the item-item graph structure by varying the number of hops $L_i$ and the $k$ value. Figure~\ref{fig: RQ6-2} shows the results of FastMMRec under different hops $L_i$ and the $k$ value on the Baby, Sports, and Clothing datasets. The suggested hops $L_i$ are 2, 1, and 1 for the Baby, Sports, and Clothing datasets, respectively. The suggested $k$ values are 10, 10, and 5 for the Baby, Sports, and Clothing datasets, respectively.
\end{itemize}
\section{Related Work}
Due to page limitations, we review recent works and their contributions in Appendix~\ref{appendix: related work}. 

\section{Testing Phase Efficiency}
Due to page limits, we provide an efficiency analysis of FastMMec in the testing phase in Appendix~\ref{appendix: inference} in supplementary materials.
\section{Conclusion}
In this work, we reveal the inevitable challenges associated with employing GCNs during the training phase in MMRec. We propose a surprisingly efficient multimodal recommendation framework for adopting graph convolution in the testing phase (FastMMRec). We conduct extensive experiments on three public datasets, consistently demonstrating the effective and efficient superiority of FastMMRec over competitive baselines. This work not only provides novel and powerful paradigms but also pinpoints potentially new research directions for efficient and large-scale real-world MMRec.

\begin{acks}
This work was supported by the Hong Kong UGC General Research Fund no. 17203320 and 17209822, and the project grants from the HKU-SCF FinTech Academy.
\end{acks}

\bibliographystyle{ACM-Reference-Format}


\clearpage
\appendix
\section{Appendix}

\begin{table*}
    \centering
    \caption{Similarity $S$ between visual and textual embeddings.}
     \vskip -0.15in
    \begin{tabular}{c|ccccccccc}
    \hline
         Models&  DualGNN&  LATTICE&  FREEDOM&  SLMRec&  BM3&  MMSSL&  LGMRec&  DiffMM& FastMMRec\\ \hline
         Baby&  0.2007&  0.2020&  0.2731&  0.3403&  0.2636&  0.3519&  0.2591&  0.2482& \textbf{0.3848}\\
         Sports&  0.1923&  0.1950&  0.2909&  0.2988&  0.2594&  0.3020&  0.2818&  0.2619& \textbf{0.3399}\\
         Clothing&  0.2005&  0.2031&  0.2678&  0.3009&  0.2657&  0.3055&  0.2833&  0.2377& \textbf{0.3400}\\ \hline
    \end{tabular}
 
    \label{tab: App2}
\end{table*}

\begin{table*}
\caption{Performance comparison of baselines on different datasets in terms of Time (s) and Memory (GB).}
 \vskip -0.15in
\centering
\label{tab: App3}
    \begin{tabular}{ccclcclcc}
    \hline
         \multirow{2}{*}{Baseline} &  \multicolumn{2}{c}{Baby}&& \multicolumn{2}{c}{Sports}&&  \multicolumn{2}{c}{Clothing}\\\cline{2-3} \cline{5-6} \cline{8-9}  
         & Time (s)& Memory (GB)&& Time (s)& Memory (GB)&& Time (s)& Memory (GB)\\\hline
         DualGNN& 7.12$e^{-5}$s& 1.80GB&& 1.33$e^{-4}$s& 4.84GB&& 1.32$e^{-4}$s& 5.42GB\\
         LATTICE& 7.14$e^{-5}$s& 1.87GB&& 1.34$e^{-4}$s& 4.96GB&& 1.37$e^{-4}$s& 5.58GB\\
         FREEDOM& 7.12$e^{-5}$s& 1.83GB&& 1.33$e^{-4}$s& 4.88GB&& 1.32$e^{-4}$s& 5.52GB\\
         SLMRec& 7.12$e^{-5}$s& 1.80GB&& 1.33$e^{-4}$s& 4.84GB&& 1.32$e^{-4}$s& 5.42GB\\
         BM3& 7.12$e^{-5}$s& 1.80GB&& 1.33$e^{-4}$s& 4.84GB&& 1.32$e^{-4}$s& 5.42GB\\
         MMSSL& 7.21$e^{-5}$s& 1.85GB&& 1.39$e^{-4}$s& 4.92GB&& 1.40$e^{-4}$s& 5.56GB\\
         LGMRec& 7.25$e^{-5}$s& 1.89GB&& 1.41$e^{-4}$s& 5.03GB&& 1.43$e^{-4}$s& 5.64GB\\
         DiffMM& 7.27$e^{-5}$s& 1.90GB&& 1.44$e^{-4}$s& 5.05GB&& 1.47$e^{-4}$s& 5.67GB\\
         FastMMRec& 7.17$e^{-5}$s& 1.80GB&& 1.37$e^{-4}$s& 4.84GB&& 1.37$e^{-4}$s& 5.42GB\\\hline
    \end{tabular}
\end{table*}

\subsection{Baseline}
\label{appendix: baseline}
In this section, we provide detailed introductions to all baseline models.
\noindent 1) Conventional recommendation models:
\begin{itemize}[leftmargin=*]
\item \textbf{MF-BPR} \cite{rendle2012bpr} leverages BPR loss to optimize the traditional collaborative filtering approach by learning representations of users and items through matrix factorization.
\item \textbf{LightGCN} \cite{he2020lightgcn} streamlines the graph convolutional network (GCN) components unnecessary for collaborative filtering, enhancing its suitability for recommendations.
\item \textbf{SimGCL} \cite{yu2022graph} proposes a graph contrastive learning that incorporates random noise directly into the feature representations.
\item \textbf{LayerGCN} \cite{zhou2023layer} employs residual connections to construct a layer-refined GCN, addressing the over-smoothing problem.
\end{itemize}

\noindent 2) Multimodal recommendation models:
\begin{itemize}[leftmargin=*]
\item \textbf{VBPR} \cite{he2016vbpr} combines visual and textual features with ID embeddings as side information for each item, effectively achieving multimodal matrix factorization.
\item \textbf{MMGCN} \cite{wei2019mmgcn} applies a GCN for each modality to learn modality-specific features and then integrates all user-predicted ratings across modalities to produce the final rating.
\item \textbf{DualGNN} \cite{wang2021dualgnn} introduces a user-user graph to uncover hidden preference patterns among users.
\item \textbf{LATTICE} \cite{zhang2021mining} develops an item-item graph to detect semantically correlated signals among items.
\item \textbf{FREEDOM} \cite{zhou2023tale} refines LATTICE by freezing the item-item graph and reducing noise in the user-item graph.
\item \textbf{SLMRec} \cite{tao2022self} proposes a self-supervised learning framework for multimodal recommendations, establishing a node self-discrimination task to reveal hidden multimodal patterns of items.
\item \textbf{BM3} \cite{zhou2023bootstrap} simplifies SLMRec by replacing the random negative example sampling mechanism with a dropout strategy.
\item \textbf{MMSSL} \cite{wei2023multi} designs a modality-aware interactive structure learning paradigm via adversarial perturbations, and proposes a cross-modal comparative learning method to disentangle the common and specific features among modalities.
\item \textbf{LGMRec} \cite{guo2024lgmrec} integrates local embeddings, which capture local topological nuances, with global embeddings, which consider hypergraph dependencies.
\item \textbf{DiffMM} \cite{jiang2024diffmm}: This method introduces a well-designed modality-aware graph diffusion model to improve modality-aware user representation learning.
\end{itemize}
\begin{table}[!ht]
\caption{Performance comparison of different strategies on all datasets in terms of NDCG@20 (N@20) and s/Epoch (\#T).}
 \vskip -0.15in
 \small
\centering
\label{tab: App1}
\setlength{\tabcolsep}{1.55mm}
\resizebox{\linewidth}{!}{
    \begin{tabular}{ccclcclcc}
    \hline
         \multirow{2}{*}{Baseline} &  \multicolumn{2}{c}{Baby}&& \multicolumn{2}{c}{Sports}&&  \multicolumn{2}{c}{Clothing}\\\cline{2-3} \cline{5-6} \cline{8-9}  
         & N@20& \#T && N@20& \#T && N@20& \#T\\\hline
         DualGNN$_{train}$& 0.0309& 5.63&& 0.0385& 11.59&& 0.0299& 14.19\\
         DualGNN$_{test}$& \textbf{0.0323}& \textbf{3.57}&& \textbf{0.0410}& \textbf{7.57}&& \textbf{0.0308}& \textbf{8.82}\\\hline
         LATTICE$_{train}$& 0.0370& 3.20&& 0.0421& 11.07&& 0.0330& 16.53\\
         LATTICE$_{test}$& \textbf{0.0383}& \textbf{2.28}&& \textbf{0.0427}& \textbf{6.01}&& \textbf{0.0339}& \textbf{8.56}\\\hline
         FREEDOM$_{train}$& 0.0424& 2.57&& 0.0481& 5.65&& 0.0420& 6.29\\
         FREEDOM$_{test}$& \textbf{0.0430}& \textbf{1.89}&& \textbf{0.0487}& \textbf{3.79}&& \textbf{0.0429}& \textbf{4.07}\\\hline
         SLMRec$_{train}$& 0.0353& 2.07&& 0.0450& 5.39&& 0.0303& 6.02\\
         SLMRec$_{test}$& \textbf{0.0359}& \textbf{1.52}&& \textbf{0.0459}& \textbf{4.28}&& \textbf{0.0310}& \textbf{4.88}\\\hline
         BM3$_{train}$& 0.0383& 1.93&& 0.0438& 3.82&& 0.0281& 5.25\\
         BM3$_{test}$& \textbf{0.0390}& \textbf{1.48}&& \textbf{0.0451}& \textbf{2.99}&& \textbf{0.0287}& \textbf{4.01}\\\hline
         MMSSL$_{train}$& 0.0420& 6.31&& 0.0474& 14.67&& 0.0359& 17.04\\
         MMSSL$_{test}$& \textbf{0.0431}& \textbf{4.61}&& \textbf{0.0482}& \textbf{8.39}&& \textbf{0.0372}& \textbf{9.59}\\\hline
    \end{tabular}
    }
\end{table}

\subsection{More Experiments for Investigation}
\label{appendix: Invest}
To further support our investigation (Section~\ref{sec: Investigation}), we conduct additional experiments on other advanced MMRec models, including DualGNN, LATTICE, FREEDOM, SLMRec, BM3, and MMSSL. (LGMRec and DiffMM are excluded as they utilize different structures—Hypergraph and Diffusion models, respectively.) As shown in Table~\ref{tab: App1}, adopting GCN in the test phase enhances both the efficiency and performance of these models.

We also provide the similarity score $S$ between visual and textual embeddings for other advanced MMRec models, including DualGNN, LATTICE, FREEDOM, SLMRec, BM3, MMSSL, LGMRec, and DiffMM. As shown in Table~\ref{tab: App2}, our FastMMRec achieves the highest similarity. While SLMRec, BM3, and MMSSL employ multiple self-supervised tasks to align visual and textual modalities, their similarity remains lower than ours (even without SSL tasks) at around 0.3, due to the limitations of GCN. Furthermore, as highlighted in the survey \cite{zhou2023comprehensive}, models with low similarity scores, such as DualGNN and LATTICE, often perform better when relying on single-modal information rather than multimodal information. This further validates the effectiveness of deploying graph convolutions exclusively during the testing phase, bypassing their use in training.

\subsection{Related Work}
\label{appendix: related work}
Many recent studies incorporate multimodal information to alleviate the data sparsity problem. VBPR \cite{he2016vbpr} utilizes visual content in conjunction with matrix factorization techniques \cite{rendle2012bpr} to mitigate data sparsity issues. Subsequent studies \cite{chen2019personalized,liu2019user,yu2023multi,xu2025mdvt,xu2025survey,chen2025squeeze,xu2025cohesion,chen2025don} have further enhanced the representation of items by incorporating both visual and textual modalities, thereby further mitigating the data sparsity problem. In an evolution of traditional recommendation system architectures, MMGCN \cite{wei2019mmgcn} employs GCN to construct a bipartite graph that extracts latent information from user-item interactions. Building on this, GRCN \cite{wei2020graph} refines the approach by pruning false-positive edges, thus reducing noise within the bipartite graph. To explicitly explore commonalities in user preferences, DualGNN \cite{wang2021dualgnn} introduces an additional user co-occurrence graph. Furthermore, LATTICE \cite{zhang2021mining} implements an item semantic graph to capture latent correlative signals between items, while FREEDOM \cite{zhou2023tale} stabilizes these representations by freezing the item semantic graph. In a novel approach, MMSSL \cite{wei2023multi} and MICRO \cite{zhang2022latent} employ contrastive self-supervised learning to align modalities and collaborative signals to enhance recommendation Additionally, BM3 \cite{zhou2023bootstrap} and PromptMM \cite{wei2024promptmm} investigate inter-modal relationships to further improve recommendation accuracy and the quality of modal fusion. LGMRec \cite{guo2024lgmrec} and DiffMM \cite{jiang2024diffmm} explore the potential of hyper-graph structures and diffusion models in enhancing the effectiveness of multimodal recommendation systems, respectively. However, the complexity of model architectures and graph learning challenges are notably amplified in MMRec. Our FastMMRec model presents a viable solution by demonstrating that adopting GCNs during the testing phase not only enhances performance relative to existing methods but also significantly boosts efficiency. Our work provides a solution for deploying MMRec in large-scale real-world scenarios.

\subsection{Testing Phase Efficiency}
\label{appendix: inference}
Inference time is critical for real-world applications. We evaluate FastMMRec and advanced baselines on per-user recommendation time and overall model memory usage. In fact, FastMMRec's graph convolution in the test phase can be converted to using graph convolution to reconstruct the user and item representations after training, without having to repeat the calculation each time in the testing phase. As shown in Table~\ref{tab: App3}, most models require similar memory usage, as they primarily store user/item embeddings, except for those with specialized structures (LGMRec, MMSSL, and DiffMM), which demand more memory. In terms of inference time, models with complex structures (LGMRec, MMSSL, and DiffMM) exhibit higher costs. FastMMRec and FREEDOM incur approximately 1\% extra inference time compared to the fastest baseline (SLMRec).

\end{document}